\documentclass[final,5p,twocolumn]{elsarticle}

\usepackage{amsmath}
\usepackage{amssymb}
\usepackage{bm}

\usepackage{graphicx}
\usepackage{booktabs}   
\usepackage{dcolumn}

\usepackage[version=4]{mhchem}

\usepackage{xcolor}
\usepackage[normalem]{ulem}

\usepackage{hyperref}
\hypersetup{
	colorlinks=true,
	linkcolor=blue,
	citecolor=blue,
	urlcolor=blue
}

\journal{Mechanics of Materials}

\usepackage{dblfloatfix} 
\usepackage[section]{placeins}

\begin{document}
	
	\begin{frontmatter}
		
		\title{Benchmarking Universal Machine Learning Interatomic Potentials for Elastic Property Prediction}
		
		\author[inst1]{Pengfei Gao}
		
		\author[inst2]{Haidi Wang\corref{cor1}}
		\ead{haidi@hfut.edu.cn}
		\cortext[cor1]{Corresponding author.}
		
		\affiliation[inst1]{organization={School of Intelligence Science and Technology, Nanjing University of Science and Technology},
			addressline={},
			city={Jiangyin},
			state={Jiangsu},
			postcode={214443},
			country={China}}
		
		\affiliation[inst2]{organization={School of Physics, Hefei University of Technology},
			addressline={},
			city={Hefei},
			state={Anhui},
			postcode={230009},
			country={China}}
		
		\begin{abstract}
			Universal machine learning interatomic potentials (uMLIPs) have emerged as efficient tools for materials simulation, yet their reliability for elastic property prediction remains unclear. Here, we present a systematic benchmark of four uMLIPs---MatterSim, MACE, SevenNet, and CHGNet---against first-principles data for nearly $11\,000$ elastically stable materials from the Materials Project database. The results show that SevenNet achieves the highest accuracy, MACE and MatterSim balance accuracy with efficiency, while CHGNet performs less effectively overall. To further improve predictive quality, we perform targeted fine-tuning on all four uMLIPs using strained configurations derived from $185$ high-error materials. After fine-tuning, CHGNet shows the most substantial improvement in overall accuracy, with MatterSim and SevenNet also benefiting from the fine-tuning, whereas MACE shows limited robustness to this procedure. This work provides quantitative guidance for model selection and data refinement, advancing uMLIPs toward reliable applications in mechanical property prediction.
		\end{abstract}
		
		
	\end{frontmatter}

	\section{Introduction}
	
	Elastic properties~\cite{10.1115/1.3423687}, as fundamental properties of materials, play an important role in governing their mechanical behavior across a wide range of applications, from structural engineering to lithium battery systems~\cite{crystal3,Kim_2021,XIE2021104602}, and related fields~\cite{phonons}. Accurate prediction of elastic constants and their derived mechanical parameters, such as bulk modulus, shear modulus, Young's modulus, and Poisson's ratio, is a critical task of computational materials design~\cite{YU2010671}. Although modern density functional theory (DFT)~\cite{hafner2006toward,hohenberg1964inhomogeneous,kohn1965self,curtarolo2005accuracy} provides reliable and reproducible predictions of elastic properties, such calculations are often associated with high computational costs in high-throughput materials screening. Owing to this computational bottleneck, the systematic exploration of large chemical spaces is strictly constrained~\cite{de2015charting}, which in turn hinders the efficient evaluation of elastic mechanical properties and delays materials design and discovery.
	
	In recent years, machine learning interatomic potentials (MLIPs)~\cite{mueller2020machine,friederich2021machine,gokcan2022learning,deringer2019machine,botu2017machine} have rapidly emerged as important tools in materials simulation, offering an effective balance between the high accuracy of quantum mechanical calculations and the efficiency of classical potentials. Generally, these models are obtained by learning interatomic interactions from large-scale DFT data sets, and enable predictions with near-quantum accuracy while substantially reducing computational cost for crystal structure prediction~\cite{crystal1,crystal2,crystal3}, molecular dynamics simulation~\cite{MD1,MD2}, and related tasks~\cite{related1,related2,cost}. Recent advances in graph neural networks, message-passing architectures, and equivariant representations have greatly improved the capabilities of MLIPs. These developments have led to the emergence of universal machine learning interatomic potentials (uMLIPs)~\cite{doi:10.1021/acsami.4c03815,chen2019graph,chen2022universal,deng2023chgnet}, which can accurately model a wide range of chemical compositions and crystal structures. However, accurately predicting elastic properties requires a dependable evaluation of the second derivatives of the potential energy surface (PES), which introduces stricter and qualitatively different challenges than those encountered in predicting energies and forces.
	
	So far, many efforts have been devoted to developing uMLIPs that improve the accuracy of energy, force, and stress predictions~\cite{deng2025systematic}. For instance, previous studies have shown that uMLIPs on the Matbench platform~\cite{dunn2020benchmarking} perform well in structural optimization, structure prediction, and molecular dynamics simulations. However, their reliability and effectiveness in predicting elastic properties remain unexplored. This is because the relationship between energy--force accuracy and second-derivative precision is not straightforward, as elastic constants are highly sensitive to slight variations in the curvature of the PES, which are often difficult to capture with conventional training strategy. Therefore, analyzing the difference between the overall predictive accuracy and property-specific performance of uMLIPs, and evaluating their capability in mechanical property predictions, is of great importance.
	
	In this work, we conduct a systematic evaluation to address the existing gap in crystal mechanical property research. Specifically, we employ four uMLIPs --- Crystal Hamiltonian Graph Neural Network (CHGNet)~\cite{deng2023chgnet}, MACE~\cite{batatia2022mace}, MatterSim~\cite{yang2024mattersim}, and Scalable EquiVariance-Enabled Neural Network (SevenNet)~\cite{park2024scalable} --- to calculate the elastic properties of $10\,994$ crystal structures from the Materials Project database~\cite{horton2025accelerated,jain2013commentary,de2015charting,jain2011formation,wang2021framework,aykol2018thermodynamic,ong2008li}, and systematically compare the results with the DFT reference data provided therein. We further quantify model performance differences in key indicators such as shear modulus, bulk modulus, Young's modulus, Poisson's ratio, and mechanical stability, as well as computational efficiency. Moreover, we introduce a targeted fine-tuning scheme that augments uMLIP training with strained configurations from high-error materials, enabling a direct evaluation of how incorporating non-equilibrium data affects mechanical predictive accuracy. Building on these analyses, we propose evidence-based guidelines for the suitable selection of uMLIPs and demonstrate that targeted fine-tuning can further enhance their accuracy and reliability in predicting elastic properties.
	
	\section{Methodology}
	
	\subsection{Dataset Construction and Analysis}
	
	\begin{figure*}[t]
		\centering
		\includegraphics[width=\textwidth]{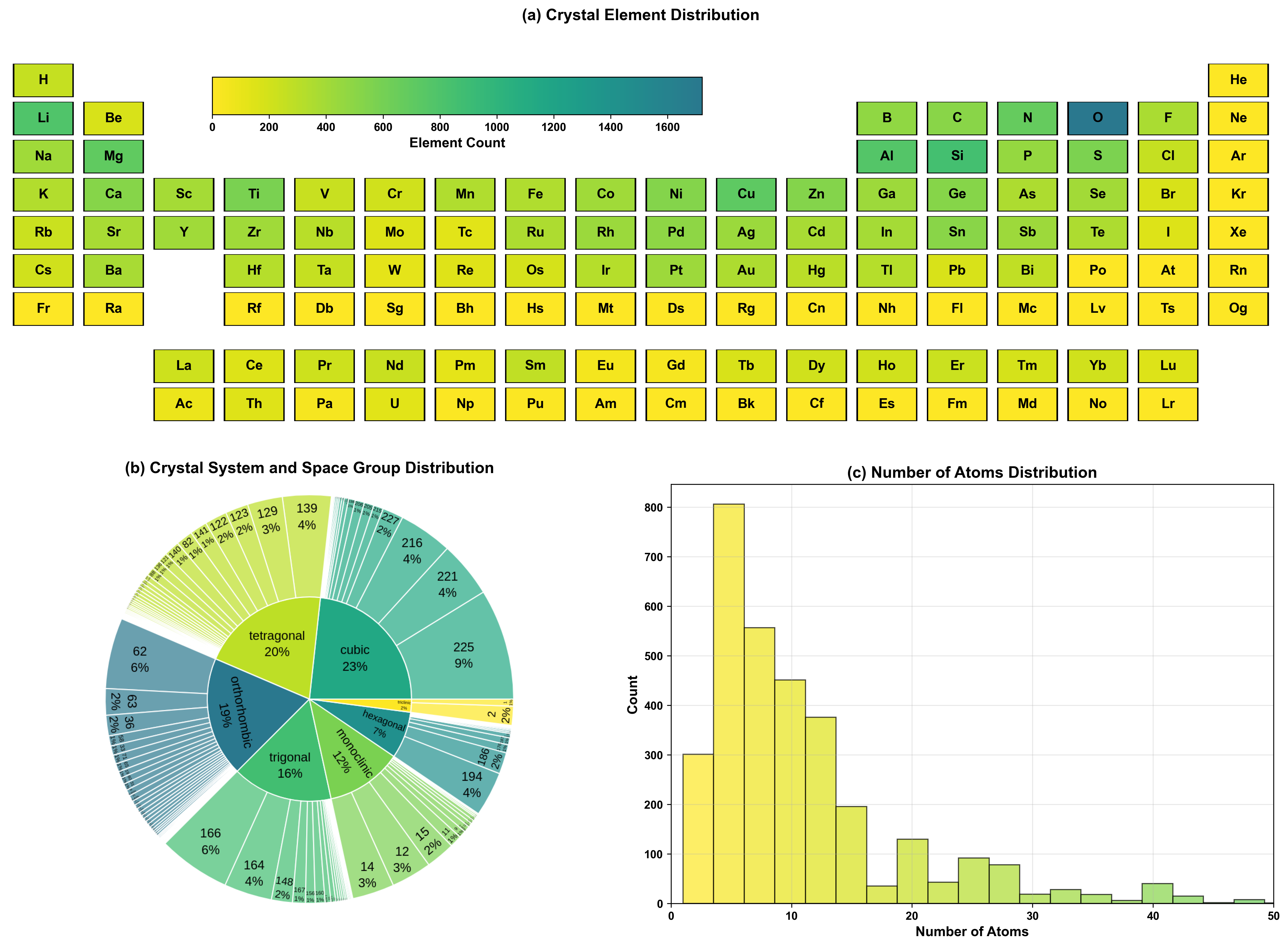}
		\caption{Crystal structure analysis of the dataset. (a) Periodic table heatmap indicating element occurrence. (b) The distribution of crystal systems and space groups, with integers placed at the outer margins representing the corresponding space-group numbers. (c) Histogram of atom counts per unit cell.}
		\label{fig:structure_analysis}
	\end{figure*}
	
	In this work, we collected $10\,994$ structures with reported elastic properties from the Materials Project database. Among them, $10\,871$ structures were mechanically stable at the DFT level, and these were used as our benchmark dataset. In Fig.~\ref{fig:structure_analysis}(a), we present the distribution of elements. The nonmetals such as \ce{B}, \ce{C}, \ce{N}, and \ce{O}, main-group metals like \ce{Li} and \ce{Mg}, and transition metals including \ce{Ni}, \ce{Cu}, \ce{Zn}, and \ce{Ti} appear most often. Heavy and radioactive elements are rare. From a crystallographic perspective (see Fig.~\ref{fig:structure_analysis}(b)), the dataset covers seven crystal systems. Cubic structures are the most common (23\%), followed by tetragonal (20\%) and orthorhombic (19\%). Trigonal and monoclinic systems make up 16\% and 12\%, while hexagonal (7\%) and triclinic (3\%) systems are less frequent. In total, 169 space groups are represented, giving wide crystallographic diversity. Finally, from the distribution of the number of atoms in Fig.~\ref{fig:structure_analysis}(c), we find that most structures have fewer than 20 atoms per unit cell, with 5--10 atoms being the most typical, and structures with more than 30 atoms are uncommon.
	
\begin{figure*}[t]
		\centering
		\includegraphics[width=\textwidth]{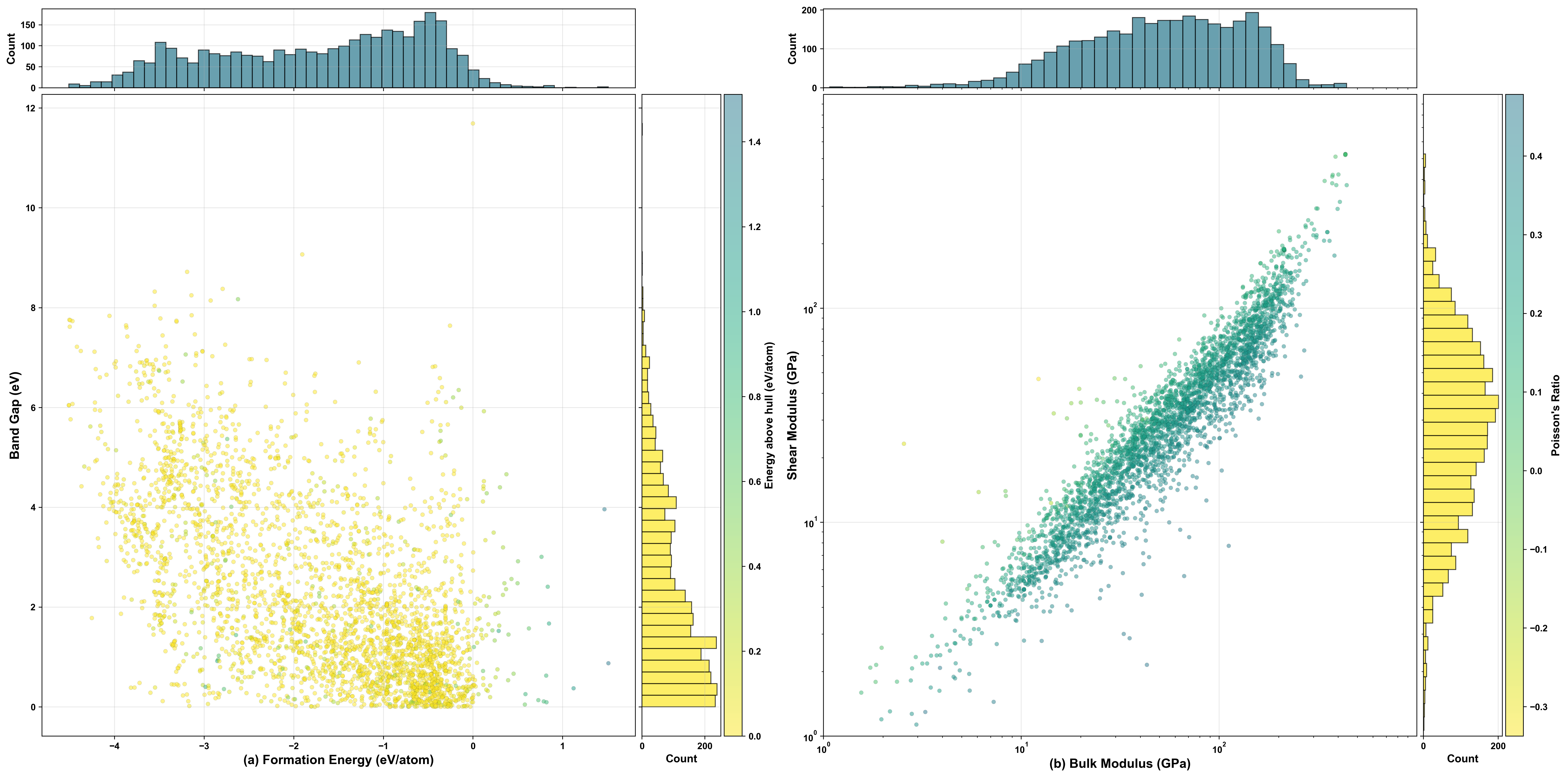}
		\caption{(a) Scatter plot of formation energy versus band gap, color-coded by the energy above hull. Marginal histograms illustrate the distribution of formation energies and band gaps. (b) Elastic property correlations. Scatter plot of bulk modulus versus shear modulus (log scale), color-coded by Poisson's ratio. Marginal histograms show the distributions of bulk and shear moduli.}
		\label{fig:electronic_elastic}
	\end{figure*}

	In Fig.~\ref{fig:electronic_elastic}, we present the distribution of the band gap, formation energy, and basic mechanical properties of the dataset, showing that these quantities exhibit a broad and diverse distribution. The statistical analysis shows that $3\,248$ materials (29.9\%) are semiconductors or insulators, with an average band gap of 0.69 eV, while the remaining $7\,623$ materials (70.1\%) are metallic. For the semiconductor subset, as illustrated in Fig.~\ref{fig:electronic_elastic}(a), the majority of structures possess negative formation energies (mean: $-0.90 \pm 0.98$ eV/atom) and energy above hull values (mean: $0.03 \pm 0.10$ eV/atom) close to zero, indicating good thermodynamic stability. Regarding mechanical properties in Fig.~\ref{fig:electronic_elastic}(b), the dataset shows that the bulk moduli range from 0.33 to 491.33 GPa (mean: $104.41 \pm 73.73$ GPa), shear moduli from 0.45 to 525.42 GPa (mean: $50.93 \pm 44.22$ GPa), and Poisson's ratios from $-0.48$ to 0.80 (mean: $0.29 \pm 0.07$). Overall, the dataset demonstrates strong representativeness in electronic, thermodynamic, and mechanical domains, providing a reliable sample for evaluating elastic properties in real materials.
	
	\subsection{Brief Description of Evaluated uMLIPs}
	
	In this work, four state-of-the-art uMLIPs were selected for comprehensive evaluation based on their elastic applications.
	
	In CHGNet~\cite{deng2023chgnet}, the total potential energy is expressed as
	\begin{equation}
		E_{\text{tot}} = \sum_i \; L_3 \circ g \circ L_2 \circ g \circ L_1 \big( \mathbf{v}_i^{(4)} \big),
	\end{equation}
	where $L_1$, $L_2$, and $L_3$ are successive linear transformations, and $g$ is a nonlinear activation function (typically the SiLU function). The vector $\mathbf{v}_i^{(4)}$ represents the final latent feature of atom $i$, obtained after four message-passing layers that aggregate both local bonding environments and longer-range structural correlations. Through this hierarchical transformation, each atomic environment is mapped to a high-dimensional representation that captures the coupling between geometric and electronic degrees of freedom. The total energy $E_{\text{tot}}$ is then constructed as a smooth, differentiable function of all atomic positions, ensuring physical consistency between predicted energies, forces, and stresses. CHGNet enhances this framework by embedding charge information into the latent space via magnetic moment constraints, which effectively incorporate electronic-structure effects into the learned potential.
	
	MACE~\cite{batatia2022mace} advances interatomic potential modeling by combining the systematic completeness of Atomic Cluster Expansion (ACE) with the higher-order equivariant message passing of modern graph neural networks. Unlike conventional message-passing neural networks that primarily encode two-body interactions and rely on deep stacking to capture higher-order correlations, MACE constructs explicit many-body messages within each layer through a hierarchical expansion,
	\begin{equation}
		\begin{split}
			m_i^{(t)} =
			& \sum_{j} u_{1}\!\big(\sigma_i^{(t)};\sigma_j^{(t)}\big)
			+\sum_{j_1,j_2} u_{2}\!\big(\sigma_i^{(t)};\sigma_{j_1}^{(t)},\sigma_{j_2}^{(t)}\big)
			+\cdots \\
			& +\sum_{j_1,\dots,j_{\nu}} u_{\nu}\!\big(\sigma_i^{(t)};\sigma_{j_1}^{(t)},\dots,\sigma_{j_{\nu}}^{(t)}\big),
		\end{split}
	\end{equation}
	where $m_i^{(t)}$ is the message received by atom $i$ at layer $t$, $\sigma_i^{(t)}=(\mathbf{r}_i, z_i, h_i^{(t)})$ denotes its geometric, chemical, and latent state, and $u_\nu$ are learnable tensorial functions encoding correlations up to body order $(\nu + 1)$.
	
	The MatterSim potential~\cite{yang2024mattersim} is a large-scale, symmetry-preserving machine-learning force field that combines the M3GNet architecture with a periodic-aware Graphormer backbone. Each atomic structure is represented as a graph $G=(Z,V,R,[L,S])$, where atomic nodes $Z$ carry feature vectors $v_i$, edges $V$ connect atom pairs $(i,j)$ within a cutoff radius $r_c$, $R=\{r_i\}$ are atomic coordinates, and $[L,S]$ encodes the global lattice and thermodynamic state. In the M3GNet message-passing block, each edge feature $e_{ij}$ represents the bond between atoms $i$ and $j$, including pairwise information such as chemical type and interatomic distance $r_{ij}=\|r_i-r_j\|$. To incorporate three-body geometry, $e_{ij}$ is refined through a spherical-Bessel / spherical-harmonic expansion of its neighboring environment:
\begin{equation}
	\begin{aligned}
		\tilde e_{ij}
		= \sum_k\;&
		j_\ell\!\Big( \tfrac{ z_{\ell n} \lVert r_{ik} \rVert }{ r_c } \Big)\,
		Y_\ell^{0}(\theta_{jik})
		\otimes \sigma(W_v v_k + b_v)
		\\
		&\qquad\times
		f_c(\lVert r_{ij} \rVert)\,
		f_c(\lVert r_{ik} \rVert)\, .
	\end{aligned}
\end{equation}

	where $\theta_{jik}$ denotes the angle between bonds $e_{ij}$ and $e_{ik}$. The smooth cutoff is
	\begin{equation}
		f_c(r) = 1 - 6(r/r_c)^5 + 15(r/r_c)^4 - 10(r/r_c)^3,
	\end{equation}
	and the updated edge feature is obtained through nonlinear mixing:
	\begin{equation}
		e'_{ij} = e_{ij} +
		g(\tilde W_2 \tilde e_{ij} + \tilde b_2)
		\otimes \sigma(\tilde W_1 \tilde e_{ij} + \tilde b_1),
	\end{equation}
	where $\sigma$ is the sigmoid activation and $g(x)=x\,\sigma(x)$.
	
	SevenNet~\cite{park2024scalable} follows the atom-decomposed energy formalism widely used in machine-learned interatomic potentials. This locality ensures that the computational cost scales linearly with the number of atoms, $\mathcal{O}(N)$, enabling large-scale molecular dynamics. At each message-passing layer $t$, atomic features are updated according to
	\begin{align}
		m_v^{(t+1)} &=
		\sum_{w\in\mathcal N(v)}
		M_t\!\big(h_v^{(t)},\, h_w^{(t)},\, e_{vw}^{(t)}\big),
		\\[4pt]
		h_v^{(t+1)} &=
		U_t\!\big(h_v^{(t)},\, m_v^{(t+1)}\big).
	\end{align}
	Here, $M_t$ and $U_t$ are learnable equivariant mappings that propagate geometric information between atoms while preserving rotational and permutational symmetries.
	
	\subsection{Elastic Property Calculations}
	
	The second-order elastic constants ($C_{ij}$) were calculated using the stress--strain method~\cite{TBarron_1965}. According to Hooke's law, the relationship between stress $\sigma_{ij}$ and strain $\varepsilon_{kl}$ with Voigt notation can be expressed as:
	\begin{equation}
		\sigma_i = C_{ij} \varepsilon_j \quad (i,j = \{1,2,3,4,5,6\}).
	\end{equation}
	The elastic tensor components are determined by applying systematic deformations to the equilibrium crystal structure and computing the resulting stress response. Based on the model-predicted stress and applied strain, the elastic constants $C_{ij}$ are obtained through linear fitting.
	
	For structure optimization and elastic-property calculations, we employ the Atomic Simulation Environment (ASE)~\cite{ase-paper,ISI:000175131400009} and Pymatgen~\cite{ong2013python}. The fast inertial relaxation engine (FIRE) algorithm~\cite{bitzek2006structural} is used for energy minimization and the FretchCellFilter~\cite{ase-paper} is applied to preserve space group symmetry during relaxation. The force convergence criterion was set to 0.1 eV/\AA{} for structure relaxation, ensuring mechanical equilibrium before strain application.
	
		The elastic tensor was calculated using a finite-strain approach as implemented in the MatCalc workflow~\cite{matcalc2024}. Once the elastic tensor is obtained, the bulk modulus, shear modulus, Young's modulus, Poisson's ratio and other derived mechanical properties can be calculated. In this work, all derived mechanical properties are reported as Voigt--Reuss--Hill (VRH) averaged values, obtained using the MechElastic~\cite{singh2021mechelastic} analysis module. The Young's modulus $E$ and Poisson's ratio $\nu$ are obtained based on the bulk modulus $K$ and shear modulus $G$ as follows:
	\begin{equation}
		E = \frac{9KG}{3K + G},
	\end{equation}
	\begin{equation}
		\nu = \frac{3K - 2G}{2(3K + G)}.
	\end{equation}
	
	\section{Results and Analysis}
	
	\subsection{Model Performance Analysis}
	
	In this section, we systematically evaluate the performance of different uMLIP models in predicting elastic properties and classifying material stability, using DFT results as the reference. By incorporating both distributional comparisons and point-wise analyses, the models are assessed from the perspectives of global trends and local accuracy. Furthermore, we also analyzed the stability classification results to provide a comprehensive picture of model applicability in elasticity-related tasks.
	
	\begin{figure*}[t]
		\centering
		\includegraphics[width=\textwidth]{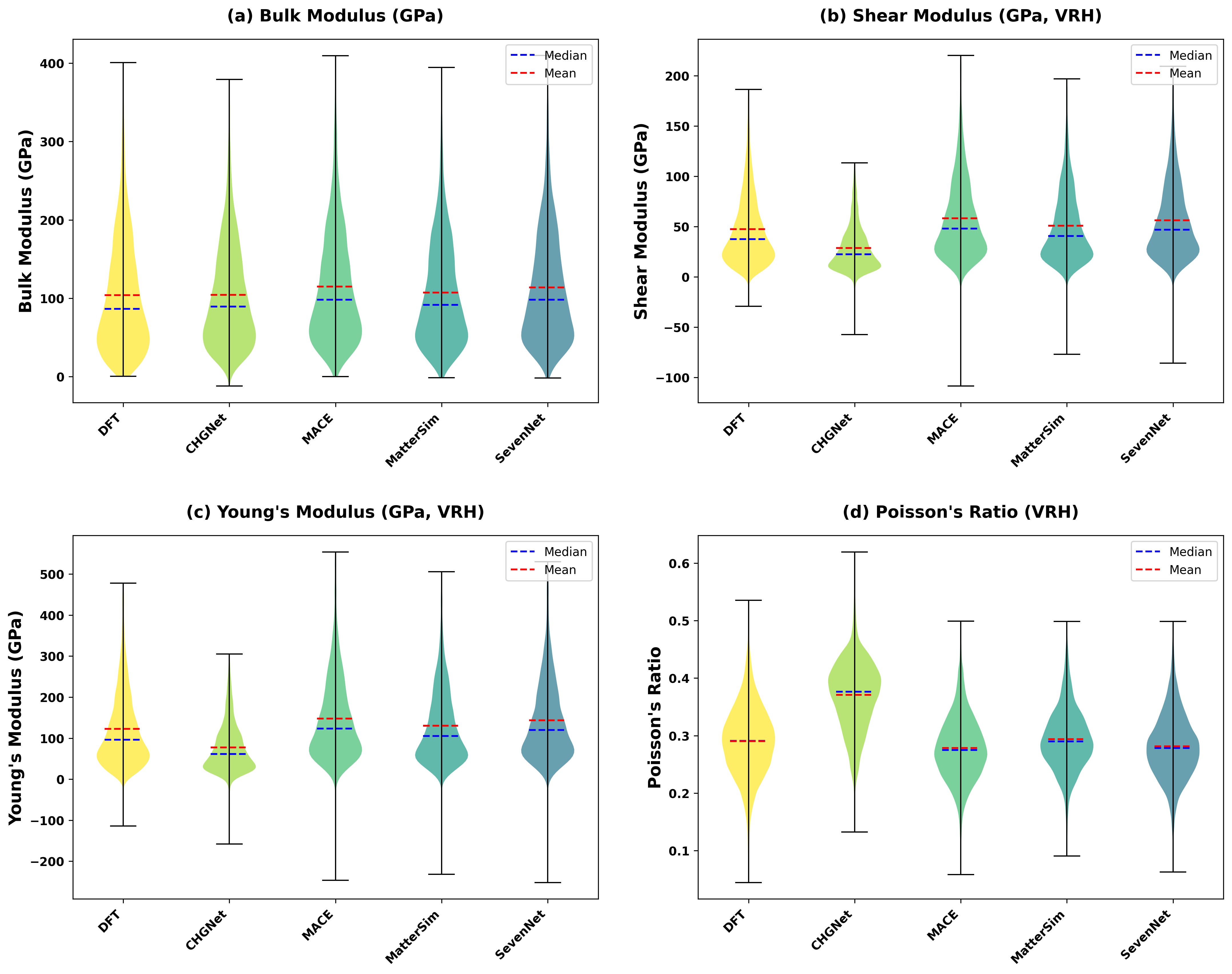}
		\caption{Distributions of (a) bulk modulus, (b) shear modulus, (c) Young's modulus, and (d) Poisson's ratio, computed as VRH averages from DFT and the four uMLIP models. Each violin plot shows the overall distribution, with blue and red dashed lines marking the median and mean, and short lines denoting the extrema.}
		\label{fig:umlip_elastic_properties}
	\end{figure*}
	
	As shown in Fig.~\ref{fig:umlip_elastic_properties}, we compare the distributions of bulk, shear, and Young's moduli and Poisson's ratio obtained from DFT and from the four uMLIP models. All models reproduce the qualitative DFT trends, but systematic and model-dependent biases persist in the absolute values. For the bulk modulus, all four models yield mean values close to the DFT reference, indicating that volumetric compressibility is robustly captured across models (DFT: 104.1 GPa; models: 100--115 GPa). In contrast, substantially larger discrepancies are observed for the shear and Young's moduli: CHGNet systematically underestimates both quantities, whereas MACE and SevenNet consistently overestimate them, with MatterSim providing intermediate values closest to DFT (DFT: 47.3/122.5 GPa; CHGNet: 28.6/77.5 GPa; MACE: 58.2/148.1 GPa; SevenNet: 56.3/143.9 GPa; MatterSim: 50.8/130.5 GPa). An opposite tendency is observed for Poisson's ratio. CHGNet markedly overestimates the DFT mean, whereas MACE and SevenNet slightly underestimate it; MatterSim again provides the closest agreement with DFT (DFT: 0.291; CHGNet: 0.371; MACE: 0.279; SevenNet: 0.282; MatterSim: 0.294).
	
	\begin{figure*}[t]
		\centering
		\includegraphics[width=\textwidth]{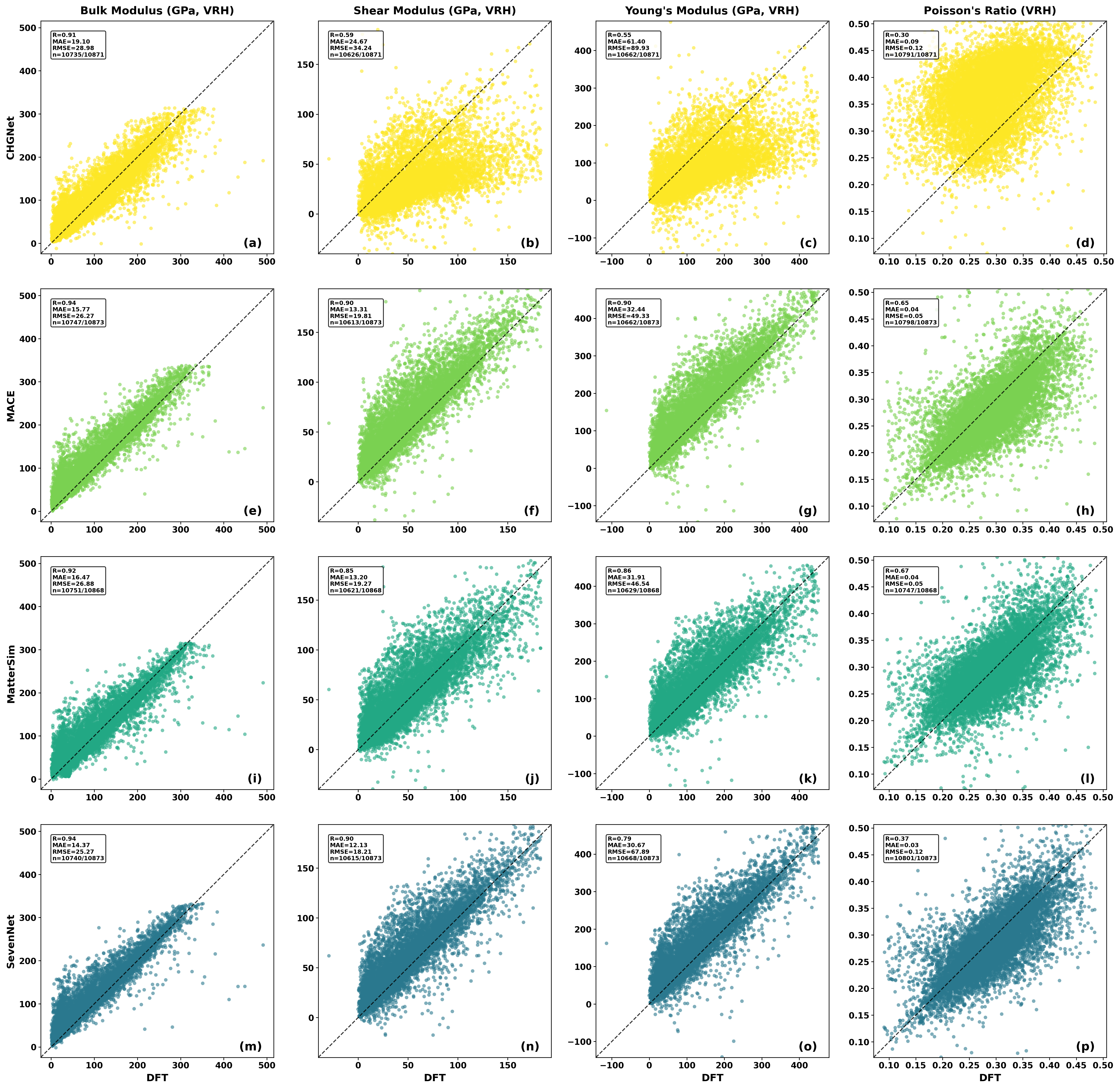}
		\caption{(a)--(h) Scatter-plot comparison of four uMLIPs against DFT reference values for primary elastic properties --- bulk modulus, shear modulus, Young's modulus, and Poisson's ratio, all computed as VRH averages. Each panel shows DFT values on the x-axis and model predictions on the y-axis, with the dashed line representing perfect agreement.}
		\label{fig:umlip_scatter_primary}
	\end{figure*}
	
	To enable quantitative evaluation, we present point-wise comparisons of the primary elastic properties in Fig.~\ref{fig:umlip_scatter_primary}. For the bulk modulus, SevenNet and MACE exhibit the highest consistency with DFT, achieving correlation coefficients of approximately $R \approx 0.94$ and mean absolute errors (MAE) around 15 GPa, outperforming both CHGNet ($R=0.909$) and MatterSim ($R=0.924$). For the shear modulus, MACE attains the highest correlation ($R=0.896$), followed by SevenNet ($R=0.895$), while MatterSim yields intermediate accuracy ($R=0.847$) and CHGNet remains significantly weaker ($R=0.546$). Regarding Young's modulus, MatterSim yields the mean closest to DFT, but in this correlation-centric assessment MACE attains the higher correlation ($R=0.901$) than MatterSim ($R=0.860$); SevenNet is lower ($R=0.791$), and CHGNet remains the weakest ($R=0.546$). For Poisson's ratio, a different trend emerges: MACE and MatterSim achieve significantly higher correlations ($R \approx 0.65$) than CHGNet ($R=0.301$) and SevenNet ($R=0.374$), indicating their robustness in capturing ratio-type properties.
	
	\begin{figure*}[t]
		\centering
		\includegraphics[width=\textwidth]{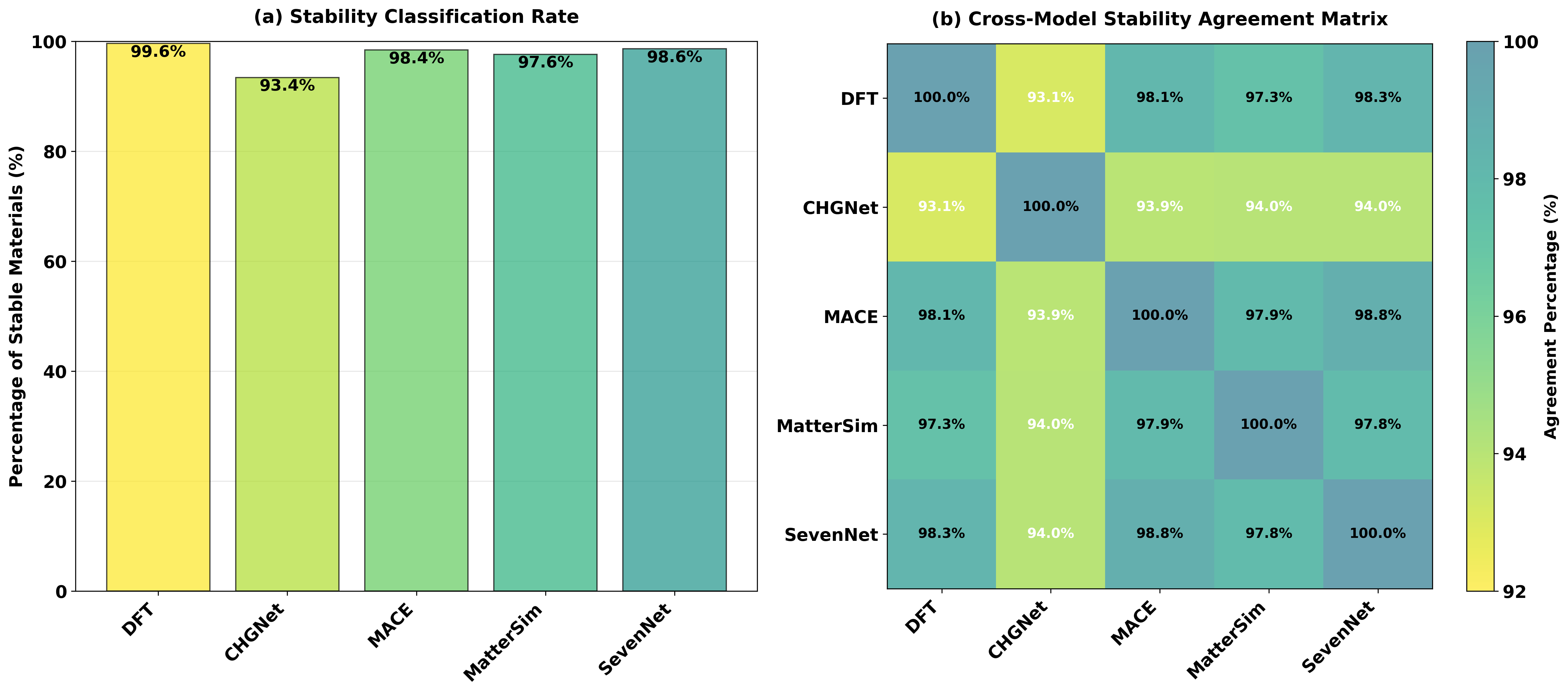}
		\caption{Elastic stability classification analysis comparing DFT and uMLIP models. (a) Stability classification rate showing the percentage of materials predicted as stable by each model. (b) Pairwise stability-agreement matrix, showing the percentage of materials assigned the same stability outcome across model pairs. Stability was determined using reported elastic stability flags or Born mechanical stability criteria.}
		\label{fig:stability_classification}
	\end{figure*}
	
	Beyond elasticity, stability classification provides another essential benchmark for evaluating model performance. Fig.~\ref{fig:stability_classification} compares the stability predictions of the four uMLIPs against DFT references. SevenNet and MACE achieve the highest performance, with accuracies of 98.3\% and 98.1\%, respectively, and F1 scores approaching 0.99, reflecting well-balanced capability in identifying both stable and unstable materials. MatterSim ranks closely behind, while CHGNet reaches only 93.4\% accuracy, significantly lower than the others and showing a higher rate of missed unstable samples.
	
	In addition, analysis of the computational efficiency for elastic property evaluations, as shown in Fig.~S1, reveals that MACE achieves the best overall performance, with an average processing time of 1.132 seconds per structure and the lowest standard deviation of 0.061 seconds. CHGNet follows closely, with an average of 1.212 seconds per structure. MatterSim has an average processing time of 1.853 seconds per structure but exhibits high standard deviation of 0.710 seconds, likely influenced by material complexity. Due to its large number of parameters, SevenNet has the highest computational cost, with an average processing time of 2.770 seconds per structure, 2.4 times that of the fastest model.
	
	\subsection{Systematic Error Analysis}
	
	To gain deeper insight into the systematic biases of different machine-learning potentials in predicting elastic properties, this section conducts a comprehensive evaluation by combining relative error distributions with mean absolute percentage error (MAPE). The joint analysis of boxplots and heatmaps reveals both the bias patterns in individual property predictions and the overall performance trends across models.
	
\begin{figure*}[t]
		\centering
		\includegraphics[width=\textwidth]{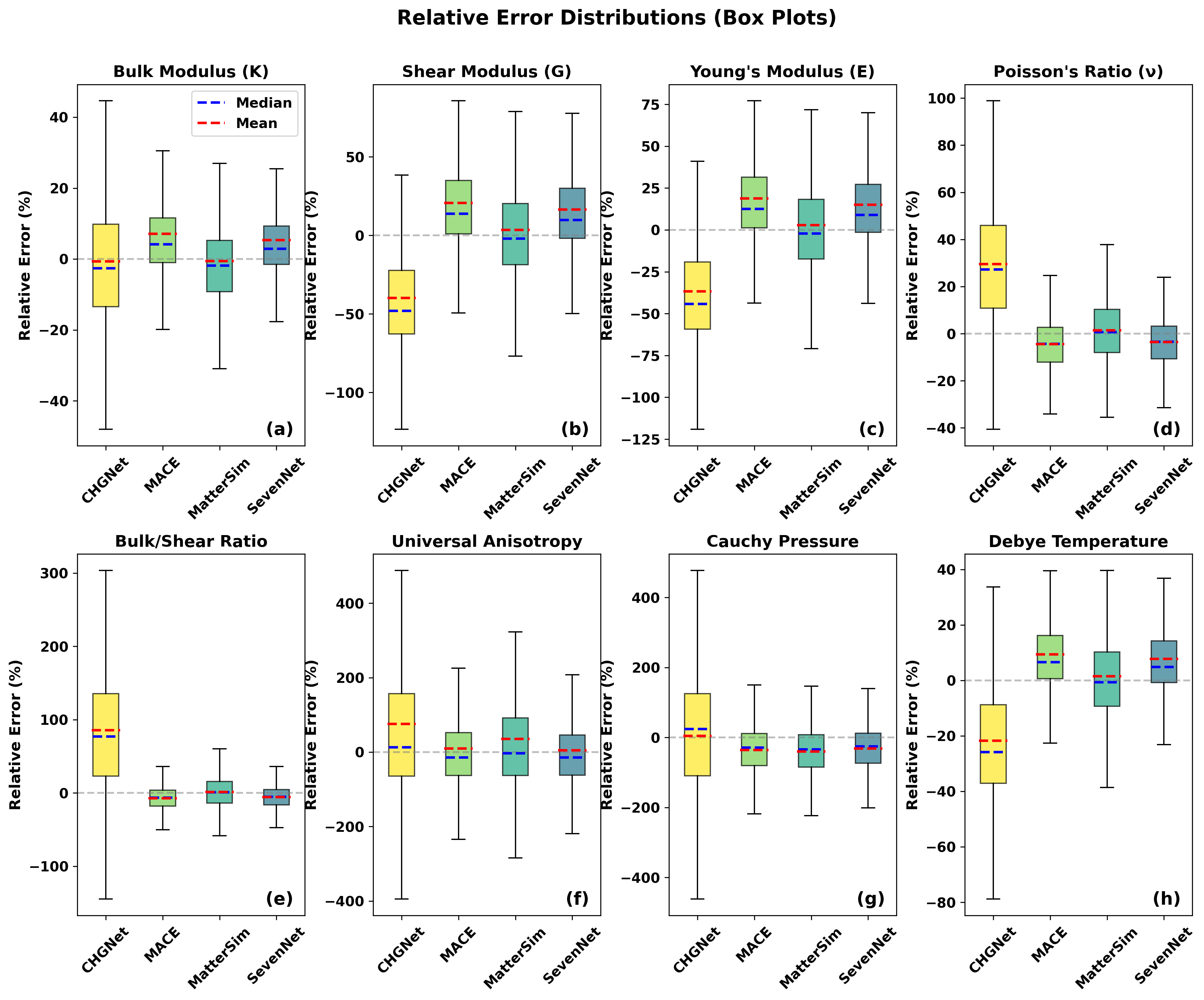}
		\caption{(a)--(h) Distribution of relative errors (\%) for CHGNet, MACE, MatterSim, and SevenNet compared with DFT reference values across eight elastic properties. Each boxplot shows the median (blue dashed line), mean (red dashed line), interquartile range (colored box), and the overall error range (short lines marking the extrema), with outliers omitted for clarity. The horizontal dashed line denotes zero error.}
		\label{fig:relative_error_boxplots}
	\end{figure*}

	Fig.~\ref{fig:relative_error_boxplots} presents the relative error distributions of the four uMLIPs with respect to DFT values across various elastic descriptors and the values in parentheses in this paragraph correspond to median relative errors. CHGNet exhibits pronounced systematic deviations across most properties. In bulk modulus predictions, it shows a median error of $-2.61\%$, indicating a tendency toward underestimation, whereas MACE and SevenNet display slight overestimations (4.16\% and 2.88\%, respectively), and MatterSim remains close to zero bias ($-1.98\%$). For the shear modulus and Young's modulus, CHGNet strongly underestimates both ($-48.02\%$ and $-44.20\%$), in sharp contrast to the overestimations observed for MACE (13.83\% and 12.43\%) and SevenNet (9.79\% and 8.89\%), while MatterSim again yields nearly symmetric distributions ($-2.12\%$ and $-2.24\%$). For Poisson's ratio, CHGNet systematically overestimates (27.25\%), opposite to the mild underestimations of MACE ($-4.35\%$) and SevenNet ($-3.40\%$), whereas MatterSim remains almost unbiased (0.70\%). The bulk/shear ratio further highlights CHGNet's strong positive bias (77.05\%), while the other models show values close to zero. CHGNet also shows especially high variability in anisotropy metrics, suggesting instability in capturing complex anisotropic behavior. For Cauchy pressure, CHGNet exhibits a systematic positive bias, whereas the others lean toward negative deviations. The large deviations in the predicted anisotropy and Cauchy pressure mainly reflect the high sensitivity of these quantities to small differences among elastic constants. In this work, most materials exhibit a relatively small degree of elastic anisotropy, with a DFT average of 1.97. The Cauchy pressure, being a difference quantity defined as $(C_{12}-C_{44})$, has a DFT average value of 17.9 GPa, which is much smaller than the bulk and Young's modulus that are generally on the order of 100 GPa. Consequently, even moderate relative errors in the stiffness components can lead to large percentage deviations in these derived quantities. Finally, in Debye temperature predictions, CHGNet again underestimates ($-25.89\%$), while MACE (6.55\%) and SevenNet (4.89\%) perform closer to DFT, and MatterSim achieves the most balanced performance ($-0.69\%$).
	
\begin{figure}[t]
		\centering
		\includegraphics[width=\columnwidth]{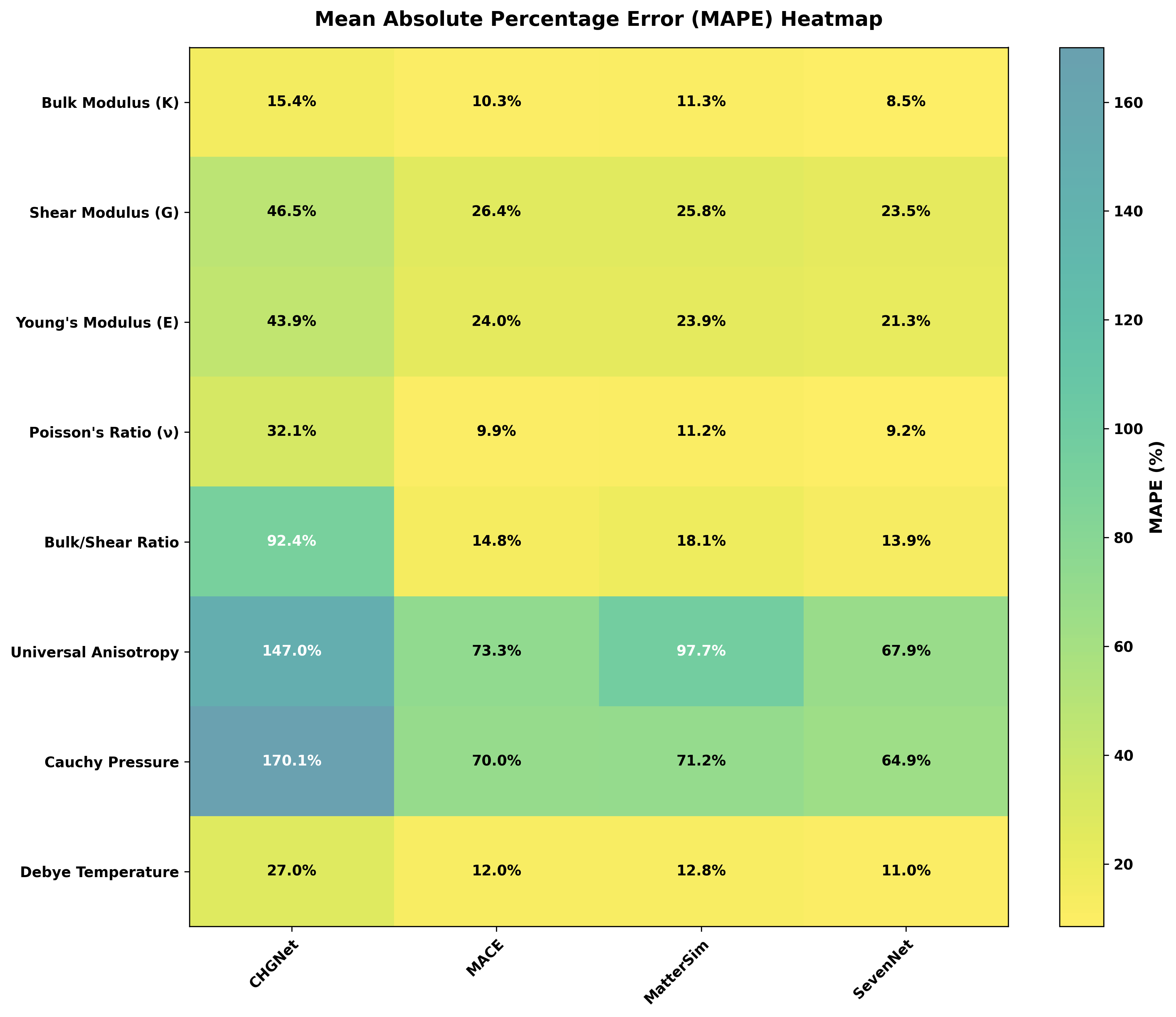}
		\caption{MAPE heatmap for elastic properties predicted by CHGNet, MACE, MatterSim, and SevenNet relative to DFT reference values. The properties analyzed include bulk modulus, shear modulus, Young's modulus, Poisson's ratio, bulk/shear ratio, universal anisotropy index, Cauchy pressure, and Debye temperature. Darker colors indicate higher errors, with values annotated in each cell.}
		\label{fig:mape_heatmap}
	\end{figure}

	To provide a clearer comparison of overall performance, Fig.~\ref{fig:mape_heatmap} summarizes the MAPE values of different models across all elastic properties. It is evident that CHGNet systematically yields the highest error levels, with an average MAPE of 71.8\%, underscoring its structural deficiencies in elastic property prediction. In contrast, SevenNet consistently achieves the lowest error, with an average MAPE of only 27.53\%, highlighting its superior overall accuracy.
	
	\subsection{Fundamental Limitations and the Impact of Fine-Tuning}
	
	The limited performance of current uMLIPs in elastic property prediction can be partially attributed to training datasets dominated by equilibrium configurations, which do not provide sufficient coverage of strained states that are important for learning accurate mechanical responses. Fine-tuning~\cite{Fine-Tuning} provides a practical route to mitigating such systematic biases. To evaluate this strategy, we constructed a targeted fine-tuning dataset comprising $185$ materials with the largest baseline errors, together with DFT-computed energies of their deformed structures, thereby explicitly introducing non-equilibrium information into the training domain. The DFT calculation settings are detailed in the Supporting Information. Fine-tuning on the targeted dataset yields well-controlled training errors for all models. The resulting mean absolute errors (MAEs) for energies, forces, and stresses are summarized in Table~\ref{tab:training_mae}.
	
	\begin{table}[t]
		\caption{Training mean absolute errors (MAEs) obtained after fine-tuning.}
		\label{tab:training_mae}
		\centering
		\begin{tabular}{lccc}
			\toprule
			Model & Energy & Force & Stress \\
			& (meV/atom) & (meV/\AA) & (meV/\AA$^3$) \\
			\midrule
			CHGNet    & 24.089 & 26.831 & 1.960 \\
			MACE      & 6.170  & 26.380 & 5.670 \\
			MatterSim & 17.890 & 9.574  & 0.836 \\
			SevenNet  & 1.692  & 5.101  & 0.528 \\
			\bottomrule
		\end{tabular}
	\end{table}
	
	All four uMLIPs --- CHGNet, MACE, MatterSim, and SevenNet --- were fine-tuned on this dataset. The updated potentials were then reassessed on the same $185$ materials. Fig.~S2 and Fig.~S3 display heatmaps of the MAPE before and after fine-tuning, while Fig.~S4 and Fig.~S5 present the corresponding distributions of relative errors. To quantify these effects, Table~\ref{tab:MAPEchange} summarizes the relative change in MAPE, defined as the difference between the fine-tuned MAPE and the original MAPE, normalized by the original value. CHGNet shows the largest improvement, with an average MAPE reduction of 23.2\%, followed by MatterSim at 20.7\% and SevenNet at 18.0\%. In contrast, MACE's average MAPE increases by 13.8\%, indicating a less favorable response to the added deformation information.
	
	\begin{table*}[t]
		\caption{Relative percentage change in the MAPE of eight elastic properties for the four uMLIPs after fine-tuning.}
		\label{tab:MAPEchange}
		\centering
		\begin{tabular}{lcccc}
			\toprule
			Property & CHGNet (\%) & MACE (\%) & MatterSim (\%) & SevenNet (\%) \\
			\midrule
			Bulk Modulus $K$           &  11.8  &   -5.2  &  -16.5  &   -2.7  \\
			Shear Modulus $G$          &   -5.3 &  -14.6  &  -15.5  &   -5.7  \\
			Young's Modulus $E$        &   -9.2 &   -9.9  &  -18.2  &   -4.5  \\
			Poisson's Ratio $\nu$      &  -56.6 &   18.4  &  -42.4  &  -48.5  \\
			Bulk/Shear Ratio           &  -55.4 &   89.4  &  -39.8  &  -41.3  \\
			Universal Anisotropy Index &  -31.6 &  37.6   &  -11.8  &  -13.9  \\
			Cauchy Pressure            &  -22.4 &   -1.0  &   -9.6  &  -18.6  \\
			Debye Temperature          &  -17.1 &   -4.3  &  -11.3  &   -9.1  \\
			\midrule
			Average                   &  -23.2 &   13.8  &  -20.7  &  -18.0  \\
			\bottomrule
		\end{tabular}
	\end{table*}
	
	\begin{table*}[t]
		\caption{IQR of the relative error for eight elastic properties before and after fine-tuning. Arrows indicate whether the IQR decreases ($\downarrow$) or increases ($\uparrow$).}
		\label{tab:IQRchange}
		\centering
		\begin{tabular}{lcccc}
			\toprule
			Property & CHGNet (\%) & MACE (\%) & MatterSim (\%) & SevenNet (\%) \\
			\midrule
			Bulk Modulus $K$
			& $236.2 \rightarrow 242.4~(\uparrow)$
			& $268.7 \rightarrow 250.4~(\downarrow)$
			& $270.6 \rightarrow 208.6~(\downarrow)$
			& $262.6 \rightarrow 260.2~(\downarrow)$ \\
			Shear Modulus $G$
			& $231.1 \rightarrow 213.7~(\downarrow)$
			& $227.1 \rightarrow 236.0~(\uparrow)$
			& $249.4 \rightarrow 193.9~(\downarrow)$
			& $232.4 \rightarrow 219.9~(\downarrow)$ \\
			Young's Modulus $E$
			& $272.7 \rightarrow 236.6~(\downarrow)$
			& $251.0 \rightarrow 258.5~(\uparrow)$
			& $267.6 \rightarrow 217.1~(\downarrow)$
			& $258.6 \rightarrow 241.9~(\downarrow)$ \\
			Poisson's Ratio $\nu$
			& $69.4 \rightarrow 29.0~(\downarrow)$
			& $36.1 \rightarrow 52.4~(\uparrow)$
			& $46.2 \rightarrow 20.6~(\downarrow)$
			& $37.5 \rightarrow 19.5~(\downarrow)$ \\
			Bulk/Shear Ratio
			& $88.2 \rightarrow 53.0~(\downarrow)$
			& $51.5 \rightarrow 104.2~(\uparrow)$
			& $60.8 \rightarrow 35.5~(\downarrow)$
			& $54.2 \rightarrow 28.8~(\downarrow)$ \\
			Universal Anisotropy Index
			& $146.4 \rightarrow 144.9~(\downarrow)$
			& $127.1 \rightarrow 162.2~(\uparrow)$
			& $116.6 \rightarrow 113.2~(\downarrow)$
			& $127.2 \rightarrow 119.7~(\downarrow)$ \\
			Cauchy Pressure
			& $174.1 \rightarrow 134.0~(\downarrow)$
			& $80.1 \rightarrow 147.4~(\uparrow)$
			& $138.2 \rightarrow 125.6~(\downarrow)$
			& $145.4 \rightarrow 116.0~(\downarrow)$ \\
			Debye Temperature
			& $99.5 \rightarrow 78.9~(\downarrow)$
			& $85.2 \rightarrow 99.9~(\uparrow)$
			& $91.4 \rightarrow 77.0~(\downarrow)$
			& $86.6 \rightarrow 81.6~(\downarrow)$ \\
			\bottomrule
		\end{tabular}
	\end{table*}
	
	\section{Discussion}
	
	\subsection{Implications for Materials Design Applications}
	
	The benchmark and fine-tuning results discussed above provide clear guidance for selecting suitable uMLIPs according to specific application requirements. For tasks requiring highly accurate predictions of elastic properties, SevenNet should be prioritized; although its computational cost is somewhat higher, it offers more reliable performance. For high-throughput screening workflows, MACE and MatterSim strike a favorable balance between accuracy and efficiency, making them better suited for large-scale applications. While CHGNet shows comparatively weaker overall performance, it remains a viable option for simulations involving magnetic systems, where its specialized capabilities can be advantageous.
	
	Systematic bias patterns observed across all models warrant careful consideration in practical applications. In particular, consistent tendencies toward underestimation or overestimation of elastic modulus highlight the need for bias-correction strategies. For quantitative materials design, it is recommended that final results be validated against high-accuracy DFT calculations to ensure reliability.
	
	\subsection{Future Directions}
	
	As demonstrated in our fine-tuning study, incorporating strained configurations into the training process can effectively improve the predictive accuracy of uMLIPs for elastic properties, underscoring the importance of dataset diversity. Future developments should focus on systematically incorporating deformed structures into training datasets, for instance through active learning strategies that aim to improve mechanical property accuracy~\cite{zhang2020dp}. For elastic property predictions within specific chemical spaces, constructing domain-specific fine-tuning~\cite{deng2025systematic} datasets and adapting pretrained models accordingly could effectively mitigate systematic biases in those regions.
	
	Improving computational efficiency remains essential for the broader adoption of uMLIPs in materials design workflows. Although current models deliver significant speedups compared to DFT, computational demands remain high when scaling to datasets containing hundreds of thousands of materials. Further optimization is therefore critical. Future developments should focus on developing hybrid frameworks that couple large-scale, low-cost screening with targeted high-accuracy calculations to ensure both efficiency and reliability in practical applications.
	
	\section{Conclusions}
	
	Our benchmark study establishes the first systematic evaluation framework for applying uMLIPs to elastic property prediction, validated across nearly $11\,000$ crystalline materials. The results demonstrate clear differences in model suitability: SevenNet delivers the highest overall accuracy, MatterSim and MACE achieve a favorable balance between accuracy and computational efficiency, while CHGNet, constrained by its architectural design, performs relatively less effectively. These results provide quantitative, evidence-based guidance for selecting suitable uMLIPs in mechanical property calculations, ensuring that model performance is properly matched to application requirements.
	
	Comprehensive analyses of systematic biases further reveal common limitations among current uMLIPs, including the consistent under- or overestimation of elastic modulus and a training-data bias toward equilibrium configurations. Our fine-tuning results, incorporating strained configurations into model training, show that even limited data augmentation can effectively mitigate such biases and enhance model robustness. CHGNet, MatterSim, and SevenNet show uniform improvements across nearly all mechanical quantities, indicating coherent and physically consistent fine-tuning behavior, whereas MACE displays smaller and more variable gains. Building on these insights, we identify several promising directions for future development, such as incorporating strained structures through active learning, implementing property-specific fine-tuning protocols, and establishing systematic error-correction schemes. We anticipate that such advances will further improve the reliability of uMLIPs for quantitative and high-throughput materials design, while also laying the groundwork for the next generation of universal interatomic potentials.
	
	
	\section*{Acknowledgements}
	This work was supported by the National Natural Science Foundation of China (22203026, 22303040) and the Fundamental Research Funds for the Central Universities (JZ2024HGTB0162). We also acknowledge the Materials Project team for providing comprehensive elastic property data and maintaining the open-access materials database.

	We acknowledge the Materials Project team for providing comprehensive elastic property data and maintaining the open-access materials database. Parts of the workflow for structure relaxation and elastic property calculations utilized or were adapted from the open-source MatCalc code developed by the Materials Project and Materials Virtual Lab.

	\section*{Author Contributions}
	Pengfei Gao performed the calculations, analyzed the results, and contributed to manuscript preparation. Haidi Wang conceived the study, performed the calculations, analyzed the results, and wrote the manuscript. All authors contributed to discussions and approved the final version of the paper.
	
	\section*{Competing Interests}
	The authors declare no competing financial or non-financial interests.
	
	\section*{Data Availability}
	All $10\,994$ crystal structures and the corresponding DFT-computed elastic property data used in this study are publicly available through the Materials Project database (\url{https://materialsproject.org}).
	
	\section*{Code Availability}
	All analysis code, model evaluation scripts, and data processing workflows are freely available at \url{https://gitee.com/haidi-hfut/umlip-elastic}. The repository includes the parameter files for the four evaluated uMLIPs and the usage instructions for the scripts for reproducing the analyses and figures.
	
	\section*{Declaration of generative AI and AI-assisted technologies in the manuscript preparation process}
During manuscript preparation, we utilized GPT-5 LLM to enhance sentence structure, readability, and coherence. We also used Claude 4 LLM for optimizing the Python plotting and simulation scripts. After using these tools, the authors reviewed and edited the content as needed and take full responsibility for the content of the published article.


\clearpage

\begin{thebibliography}{47}
	\expandafter\ifx\csname natexlab\endcsname\relax\def\natexlab#1{#1}\fi
	\providecommand{\url}[1]{\texttt{#1}}
	\providecommand{\href}[2]{#2}
	\providecommand{\path}[1]{#1}
	\providecommand{\DOIprefix}{doi:}
	\providecommand{\ArXivprefix}{arXiv:}
	\providecommand{\URLprefix}{URL: }
	\providecommand{\Pubmedprefix}{pmid:}
	\providecommand{\doi}[1]{\href{http://dx.doi.org/#1}{\path{#1}}}
	\providecommand{\Pubmed}[1]{\href{pmid:#1}{\path{#1}}}
	\providecommand{\bibinfo}[2]{#2}
	\ifx\xfnm\relax \def\xfnm[#1]{\unskip,\space#1}\fi
	\bibitem[{Schreiber et~al.(1975)Schreiber, Anderson, Soga, and
		Bell}]{10.1115/1.3423687}
	\bibinfo{author}{E.~Schreiber}, \bibinfo{author}{O.~L. Anderson},
	\bibinfo{author}{N.~Soga}, \bibinfo{author}{J.~F. Bell},
	\newblock \bibinfo{title}{Elastic constants and their measurement},
	\newblock \bibinfo{journal}{Journal of Applied Mechanics} \bibinfo{volume}{42}
	(\bibinfo{year}{1975}) \bibinfo{pages}{747--748}.
	\bibitem[{Wu et~al.(2022)Wu, Kim, Lee, Um, Lee, Lai, Byun, and Chou}]{crystal3}
	\bibinfo{author}{C.~Wu}, \bibinfo{author}{T.~Kim}, \bibinfo{author}{S.-B. Lee},
	\bibinfo{author}{M.-K. Um}, \bibinfo{author}{S.-K. Lee},
	\bibinfo{author}{W.-Y. Lai}, \bibinfo{author}{J.-H. Byun},
	\bibinfo{author}{T.-W. Chou},
	\newblock \bibinfo{title}{An overview of composite structural engineering for
		stretchable strain sensors},
	\newblock \bibinfo{journal}{Composites Science and Technology}
	\bibinfo{volume}{229} (\bibinfo{year}{2022}).
	\bibitem[{Kim et~al.(2021)Kim, Yang, and Bloom}]{Kim_2021}
	\bibinfo{author}{M.~Kim}, \bibinfo{author}{Z.~Yang},
	\bibinfo{author}{I.~Bloom},
	\newblock \bibinfo{title}{Review—the lithiation/delithiation behavior of
		si-based electrodes: A connection between electrochemistry and mechanics},
	\newblock \bibinfo{journal}{Journal of The Electrochemical Society}
	\bibinfo{volume}{168} (\bibinfo{year}{2021}) \bibinfo{pages}{010523}.
	\bibitem[{Xie et~al.(2021)Xie, Han, Song, Li, Kang, and Zhang}]{XIE2021104602}
	\bibinfo{author}{H.~Xie}, \bibinfo{author}{B.~Han}, \bibinfo{author}{H.~Song},
	\bibinfo{author}{X.~Li}, \bibinfo{author}{Y.~Kang},
	\bibinfo{author}{Q.~Zhang},
	\newblock \bibinfo{title}{In-situ measurements of electrochemical stress/strain
		fields and stress analysis during an electrochemical process},
	\newblock \bibinfo{journal}{Journal of the Mechanics and Physics of Solids}
	\bibinfo{volume}{156} (\bibinfo{year}{2021}) \bibinfo{pages}{104602}.
	\bibitem[{Loew et~al.(2025)Loew, Sun, Wang, Botti, and Marques}]{phonons}
	\bibinfo{author}{A.~Loew}, \bibinfo{author}{D.~Sun}, \bibinfo{author}{H.-C.
		Wang}, \bibinfo{author}{S.~Botti}, \bibinfo{author}{M.~A. Marques},
	\newblock \bibinfo{title}{Universal machine learning interatomic potentials are
		ready for phonons},
	\newblock \bibinfo{journal}{npj Computational Materials} \bibinfo{volume}{11}
	(\bibinfo{year}{2025}) \bibinfo{pages}{178}.
	\bibitem[{Yu et~al.(2010)Yu, Zhu, and Ye}]{YU2010671}
	\bibinfo{author}{R.~Yu}, \bibinfo{author}{J.~Zhu}, \bibinfo{author}{H.~Ye},
	\newblock \bibinfo{title}{Calculations of single-crystal elastic constants made
		simple},
	\newblock \bibinfo{journal}{Computer Physics Communications}
	\bibinfo{volume}{181} (\bibinfo{year}{2010}) \bibinfo{pages}{671--675}.
	\bibitem[{Hafner et~al.(2006)Hafner, Wolverton, and Ceder}]{hafner2006toward}
	\bibinfo{author}{J.~Hafner}, \bibinfo{author}{C.~Wolverton},
	\bibinfo{author}{G.~Ceder},
	\newblock \bibinfo{title}{Toward computational materials design: the impact of
		density functional theory on materials research},
	\newblock \bibinfo{journal}{MRS Bulletin} \bibinfo{volume}{31}
	(\bibinfo{year}{2006}) \bibinfo{pages}{659--668}.
	\bibitem[{Hohenberg and Kohn(1964)}]{hohenberg1964inhomogeneous}
	\bibinfo{author}{P.~Hohenberg}, \bibinfo{author}{W.~Kohn},
	\newblock \bibinfo{title}{Inhomogeneous electron gas},
	\newblock \bibinfo{journal}{Physical Review} \bibinfo{volume}{136}
	(\bibinfo{year}{1964}) \bibinfo{pages}{B864}.
	\bibitem[{Kohn and Sham(1965)}]{kohn1965self}
	\bibinfo{author}{W.~Kohn}, \bibinfo{author}{L.~J. Sham},
	\newblock \bibinfo{title}{Self-consistent equations including exchange and
		correlation effects},
	\newblock \bibinfo{journal}{Physical Review} \bibinfo{volume}{140}
	(\bibinfo{year}{1965}) \bibinfo{pages}{A1133}.
	\bibitem[{Curtarolo et~al.(2005)Curtarolo, Morgan, and
		Ceder}]{curtarolo2005accuracy}
	\bibinfo{author}{S.~Curtarolo}, \bibinfo{author}{D.~Morgan},
	\bibinfo{author}{G.~Ceder},
	\newblock \bibinfo{title}{Accuracy of ab initio methods in predicting the
		crystal structures of metals: A review of 80 binary alloys},
	\newblock \bibinfo{journal}{Calphad} \bibinfo{volume}{29}
	(\bibinfo{year}{2005}) \bibinfo{pages}{163--211}.
	\bibitem[{De~Jong et~al.(2015)De~Jong, Chen, Angsten, Jain, Notestine, Gamst,
		Sluiter, Krishna~Ande, Van Der~Zwaag, Plata et~al.}]{de2015charting}
	\bibinfo{author}{M.~De~Jong}, \bibinfo{author}{W.~Chen},
	\bibinfo{author}{T.~Angsten}, \bibinfo{author}{A.~Jain},
	\bibinfo{author}{R.~Notestine}, \bibinfo{author}{A.~Gamst},
	\bibinfo{author}{M.~Sluiter}, \bibinfo{author}{C.~Krishna~Ande},
	\bibinfo{author}{S.~Van Der~Zwaag}, \bibinfo{author}{J.~J. Plata}, et~al.,
	\newblock \bibinfo{title}{Charting the complete elastic properties of inorganic
		crystalline compounds},
	\newblock \bibinfo{journal}{Scientific Data} \bibinfo{volume}{2}
	(\bibinfo{year}{2015}) \bibinfo{pages}{1--13}.
	\bibitem[{Mueller et~al.(2020)Mueller, Hernandez, and
		Wang}]{mueller2020machine}
	\bibinfo{author}{T.~Mueller}, \bibinfo{author}{A.~Hernandez},
	\bibinfo{author}{C.~Wang},
	\newblock \bibinfo{title}{Machine learning for interatomic potential models},
	\newblock \bibinfo{journal}{The Journal of Chemical Physics}
	\bibinfo{volume}{152} (\bibinfo{year}{2020}) \bibinfo{pages}{050902}.
	\bibitem[{Friederich et~al.(2021)Friederich, H{\"a}se, Proppe, and
		Aspuru-Guzik}]{friederich2021machine}
	\bibinfo{author}{P.~Friederich}, \bibinfo{author}{F.~H{\"a}se},
	\bibinfo{author}{J.~Proppe}, \bibinfo{author}{A.~Aspuru-Guzik},
	\newblock \bibinfo{title}{Machine-learned potentials for next-generation matter
		simulations},
	\newblock \bibinfo{journal}{Nature Materials} \bibinfo{volume}{20}
	(\bibinfo{year}{2021}) \bibinfo{pages}{750--761}.
	\bibitem[{Gokcan and Isayev(2022)}]{gokcan2022learning}
	\bibinfo{author}{H.~Gokcan}, \bibinfo{author}{O.~Isayev},
	\newblock \bibinfo{title}{Learning molecular potentials with neural networks},
	\newblock \bibinfo{journal}{Wiley Interdisciplinary Reviews: Computational
		Molecular Science} \bibinfo{volume}{12} (\bibinfo{year}{2022})
	\bibinfo{pages}{e1564}.
	\bibitem[{Deringer et~al.(2019)Deringer, Caro, and
		Cs{\'a}nyi}]{deringer2019machine}
	\bibinfo{author}{V.~L. Deringer}, \bibinfo{author}{M.~A. Caro},
	\bibinfo{author}{G.~Cs{\'a}nyi},
	\newblock \bibinfo{title}{Machine learning interatomic potentials as emerging
		tools for materials science},
	\newblock \bibinfo{journal}{Advanced Materials} \bibinfo{volume}{31}
	(\bibinfo{year}{2019}) \bibinfo{pages}{1902765}.
	\bibitem[{Botu et~al.(2017)Botu, Batra, Chapman, and
		Ramprasad}]{botu2017machine}
	\bibinfo{author}{V.~Botu}, \bibinfo{author}{R.~Batra},
	\bibinfo{author}{J.~Chapman}, \bibinfo{author}{R.~Ramprasad},
	\newblock \bibinfo{title}{Machine learning force fields: construction,
		validation, and outlook},
	\newblock \bibinfo{journal}{The Journal of Physical Chemistry C}
	\bibinfo{volume}{121} (\bibinfo{year}{2017}) \bibinfo{pages}{511--522}.
	\bibitem[{Podryabinkin et~al.(2019)Podryabinkin, Tikhonov, Shapeev, and
		Oganov}]{crystal1}
	\bibinfo{author}{E.~V. Podryabinkin}, \bibinfo{author}{E.~V. Tikhonov},
	\bibinfo{author}{A.~V. Shapeev}, \bibinfo{author}{A.~R. Oganov},
	\newblock \bibinfo{title}{Accelerating crystal structure prediction by
		machine-learning interatomic potentials with active learning},
	\newblock \bibinfo{journal}{Physical Review B} \bibinfo{volume}{99}
	(\bibinfo{year}{2019}) \bibinfo{pages}{064114}.
	\bibitem[{Paleico and Behler(2020)}]{crystal2}
	\bibinfo{author}{M.~L. Paleico}, \bibinfo{author}{J.~Behler},
	\newblock \bibinfo{title}{Global optimization of copper clusters at the {ZnO
			(101 0)} surface using a {DFT-based} neural network potential and genetic
		algorithms},
	\newblock \bibinfo{journal}{The Journal of Chemical Physics}
	\bibinfo{volume}{153} (\bibinfo{year}{2020}) \bibinfo{pages}{054704}.
	\bibitem[{Xu et~al.(2023)Xu, Duan, Dou, Zheng, Lin, Xia, Zhao, and Xia}]{MD1}
	\bibinfo{author}{Z.~Xu}, \bibinfo{author}{H.~Duan}, \bibinfo{author}{Z.~Dou},
	\bibinfo{author}{M.~Zheng}, \bibinfo{author}{Y.~Lin},
	\bibinfo{author}{Y.~Xia}, \bibinfo{author}{H.~Zhao},
	\bibinfo{author}{Y.~Xia},
	\newblock \bibinfo{title}{Machine learning molecular dynamics simulation
		identifying weakly negative effect of polyanion rotation on {Li-ion}
		migration},
	\newblock \bibinfo{journal}{npj Computational Materials} \bibinfo{volume}{9}
	(\bibinfo{year}{2023}) \bibinfo{pages}{105}.
	\bibitem[{Zhang et~al.(2021)Zhang, Wang, Car, and E}]{MD2}
	\bibinfo{author}{L.~Zhang}, \bibinfo{author}{H.~Wang},
	\bibinfo{author}{R.~Car}, \bibinfo{author}{W.~E},
	\newblock \bibinfo{title}{Phase diagram of a deep potential water model},
	\newblock \bibinfo{journal}{Physical Review Letters} \bibinfo{volume}{126}
	(\bibinfo{year}{2021}) \bibinfo{pages}{236001}.
	\bibitem[{Rosenbrock et~al.(2021)Rosenbrock, Gubaev, Shapeev, P{\'a}rtay,
		Bernstein, Cs{\'a}nyi, and Hart}]{related1}
	\bibinfo{author}{C.~W. Rosenbrock}, \bibinfo{author}{K.~Gubaev},
	\bibinfo{author}{A.~V. Shapeev}, \bibinfo{author}{L.~B. P{\'a}rtay},
	\bibinfo{author}{N.~Bernstein}, \bibinfo{author}{G.~Cs{\'a}nyi},
	\bibinfo{author}{G.~L. Hart},
	\newblock \bibinfo{title}{Machine-learned interatomic potentials for alloys and
		alloy phase diagrams},
	\newblock \bibinfo{journal}{npj Computational Materials} \bibinfo{volume}{7}
	(\bibinfo{year}{2021}) \bibinfo{pages}{24}.
	\bibitem[{Kulichenko et~al.(2024)Kulichenko, Nebgen, Lubbers, Smith, Barros,
		Allen, Habib, Shinkle, Fedik, Li et~al.}]{related2}
	\bibinfo{author}{M.~Kulichenko}, \bibinfo{author}{B.~Nebgen},
	\bibinfo{author}{N.~Lubbers}, \bibinfo{author}{J.~S. Smith},
	\bibinfo{author}{K.~Barros}, \bibinfo{author}{A.~E. Allen},
	\bibinfo{author}{A.~Habib}, \bibinfo{author}{E.~Shinkle},
	\bibinfo{author}{N.~Fedik}, \bibinfo{author}{Y.~W. Li}, et~al.,
	\newblock \bibinfo{title}{Data generation for machine learning interatomic
		potentials and beyond},
	\newblock \bibinfo{journal}{Chemical Reviews} \bibinfo{volume}{124}
	(\bibinfo{year}{2024}) \bibinfo{pages}{13681--13714}.
	\bibitem[{Zuo et~al.(2020)Zuo, Chen, Li, Deng, Chen, Behler, Cs{\'a}nyi,
		Shapeev, Thompson, Wood et~al.}]{cost}
	\bibinfo{author}{Y.~Zuo}, \bibinfo{author}{C.~Chen}, \bibinfo{author}{X.~Li},
	\bibinfo{author}{Z.~Deng}, \bibinfo{author}{Y.~Chen},
	\bibinfo{author}{J.~Behler}, \bibinfo{author}{G.~Cs{\'a}nyi},
	\bibinfo{author}{A.~V. Shapeev}, \bibinfo{author}{A.~P. Thompson},
	\bibinfo{author}{M.~A. Wood}, et~al.,
	\newblock \bibinfo{title}{Performance and cost assessment of machine learning
		interatomic potentials},
	\newblock \bibinfo{journal}{The Journal of Physical Chemistry A}
	\bibinfo{volume}{124} (\bibinfo{year}{2020}) \bibinfo{pages}{731--745}.
	\bibitem[{Focassio et~al.(2025)Focassio, M.~Freitas, and
		Schleder}]{doi:10.1021/acsami.4c03815}
	\bibinfo{author}{B.~Focassio}, \bibinfo{author}{L.~P. M.~Freitas},
	\bibinfo{author}{G.~R. Schleder},
	\newblock \bibinfo{title}{Performance assessment of universal machine learning
		interatomic potentials: Challenges and directions for materials’ surfaces},
	\newblock \bibinfo{journal}{ACS Applied Materials \& Interfaces}
	\bibinfo{volume}{17} (\bibinfo{year}{2025}) \bibinfo{pages}{13111--13121}.
	\bibitem[{Chen et~al.(2019)Chen, Ye, Zuo, Zheng, and Ong}]{chen2019graph}
	\bibinfo{author}{C.~Chen}, \bibinfo{author}{W.~Ye}, \bibinfo{author}{Y.~Zuo},
	\bibinfo{author}{C.~Zheng}, \bibinfo{author}{S.~P. Ong},
	\newblock \bibinfo{title}{Graph networks as a universal machine learning
		framework for molecules and crystals},
	\newblock \bibinfo{journal}{Chemistry of Materials} \bibinfo{volume}{31}
	(\bibinfo{year}{2019}) \bibinfo{pages}{3564--3572}.
	\bibitem[{Chen and Ong(2022)}]{chen2022universal}
	\bibinfo{author}{C.~Chen}, \bibinfo{author}{S.~P. Ong},
	\newblock \bibinfo{title}{A universal graph deep learning interatomic potential
		for the periodic table},
	\newblock \bibinfo{journal}{Nature Computational Science} \bibinfo{volume}{2}
	(\bibinfo{year}{2022}) \bibinfo{pages}{718--728}.
	\bibitem[{Deng et~al.(2023)Deng, Zhong, Jun, Riebesell, Han, Bartel, and
		Ceder}]{deng2023chgnet}
	\bibinfo{author}{B.~Deng}, \bibinfo{author}{P.~Zhong},
	\bibinfo{author}{K.~Jun}, \bibinfo{author}{J.~Riebesell},
	\bibinfo{author}{K.~Han}, \bibinfo{author}{C.~J. Bartel},
	\bibinfo{author}{G.~Ceder},
	\newblock \bibinfo{title}{Chgnet as a pretrained universal neural network
		potential for charge-informed atomistic modelling},
	\newblock \bibinfo{journal}{Nature Machine Intelligence} \bibinfo{volume}{5}
	(\bibinfo{year}{2023}) \bibinfo{pages}{1031--1041}.
	\bibitem[{Deng et~al.(2025)Deng, Choi, Zhong, Riebesell, Anand, Li, Jun,
		Persson, and Ceder}]{deng2025systematic}
	\bibinfo{author}{B.~Deng}, \bibinfo{author}{Y.~Choi},
	\bibinfo{author}{P.~Zhong}, \bibinfo{author}{J.~Riebesell},
	\bibinfo{author}{S.~Anand}, \bibinfo{author}{Z.~Li},
	\bibinfo{author}{K.~Jun}, \bibinfo{author}{K.~A. Persson},
	\bibinfo{author}{G.~Ceder},
	\newblock \bibinfo{title}{Systematic softening in universal machine learning
		interatomic potentials},
	\newblock \bibinfo{journal}{npj Computational Materials} \bibinfo{volume}{11}
	(\bibinfo{year}{2025}) \bibinfo{pages}{9}.
	\bibitem[{Dunn et~al.(2020)Dunn, Wang, Ganose, Dopp, and
		Jain}]{dunn2020benchmarking}
	\bibinfo{author}{A.~Dunn}, \bibinfo{author}{Q.~Wang},
	\bibinfo{author}{A.~Ganose}, \bibinfo{author}{D.~Dopp},
	\bibinfo{author}{A.~Jain},
	\newblock \bibinfo{title}{Benchmarking materials property prediction methods:
		the matbench test set and automatminer reference algorithm},
	\newblock \bibinfo{journal}{npj Computational Materials} \bibinfo{volume}{6}
	(\bibinfo{year}{2020}) \bibinfo{pages}{138}.
	\bibitem[{Batatia et~al.(2022)Batatia, Kovacs, Simm, Ortner, and
		Cs{\'a}nyi}]{batatia2022mace}
	\bibinfo{author}{I.~Batatia}, \bibinfo{author}{D.~P. Kovacs},
	\bibinfo{author}{G.~Simm}, \bibinfo{author}{C.~Ortner},
	\bibinfo{author}{G.~Cs{\'a}nyi},
	\newblock \bibinfo{title}{Mace: Higher order equivariant message passing neural
		networks for fast and accurate force fields},
	\newblock \bibinfo{journal}{Advances in Neural Information Processing Systems}
	\bibinfo{volume}{35} (\bibinfo{year}{2022}) \bibinfo{pages}{11423--11436}.
	\bibitem[{Yang et~al.(2024)Yang, Hu, Zhou, Liu, Shi, Li, Li, Chen, Chen, Zeni
		et~al.}]{yang2024mattersim}
	\bibinfo{author}{H.~Yang}, \bibinfo{author}{C.~Hu}, \bibinfo{author}{Y.~Zhou},
	\bibinfo{author}{X.~Liu}, \bibinfo{author}{Y.~Shi}, \bibinfo{author}{J.~Li},
	\bibinfo{author}{G.~Li}, \bibinfo{author}{Z.~Chen},
	\bibinfo{author}{S.~Chen}, \bibinfo{author}{C.~Zeni}, et~al.,
	\newblock \bibinfo{title}{Mattersim: A deep learning atomistic model across
		elements, temperatures and pressures},
	\newblock \bibinfo{journal}{arXiv:2405.04967}  (\bibinfo{year}{2024}).
	\bibitem[{Park et~al.(2024)Park, Kim, Hwang, and Han}]{park2024scalable}
	\bibinfo{author}{Y.~Park}, \bibinfo{author}{J.~Kim},
	\bibinfo{author}{S.~Hwang}, \bibinfo{author}{S.~Han},
	\newblock \bibinfo{title}{Scalable parallel algorithm for graph neural network
		interatomic potentials in molecular dynamics simulations},
	\newblock \bibinfo{journal}{Journal of Chemical Theory and Computation}
	\bibinfo{volume}{20} (\bibinfo{year}{2024}) \bibinfo{pages}{4857--4868}.
	\bibitem[{Horton et~al.(2025)Horton, Huck, Yang, Munro, Dwaraknath, Ganose,
		Kingsbury, Wen, Shen, Mathis et~al.}]{horton2025accelerated}
	\bibinfo{author}{M.~K. Horton}, \bibinfo{author}{P.~Huck},
	\bibinfo{author}{R.~X. Yang}, \bibinfo{author}{J.~M. Munro},
	\bibinfo{author}{S.~Dwaraknath}, \bibinfo{author}{A.~M. Ganose},
	\bibinfo{author}{R.~S. Kingsbury}, \bibinfo{author}{M.~Wen},
	\bibinfo{author}{J.~X. Shen}, \bibinfo{author}{T.~S. Mathis}, et~al.,
	\newblock \bibinfo{title}{Accelerated data-driven materials science with the
		materials project},
	\newblock \bibinfo{journal}{Nature Materials} \bibinfo{volume}{24}
	(\bibinfo{year}{2025}) \bibinfo{pages}{1522--1532}.
	\bibitem[{Jain et~al.(2013)Jain, Ong, Hautier, Chen, Richards, Dacek, Cholia,
		Gunter, Skinner, Ceder et~al.}]{jain2013commentary}
	\bibinfo{author}{A.~Jain}, \bibinfo{author}{S.~P. Ong},
	\bibinfo{author}{G.~Hautier}, \bibinfo{author}{W.~Chen},
	\bibinfo{author}{W.~D. Richards}, \bibinfo{author}{S.~Dacek},
	\bibinfo{author}{S.~Cholia}, \bibinfo{author}{D.~Gunter},
	\bibinfo{author}{D.~Skinner}, \bibinfo{author}{G.~Ceder}, et~al.,
	\newblock \bibinfo{title}{Commentary: The materials project: A materials genome
		approach to accelerating materials innovation},
	\newblock \bibinfo{journal}{APL Materials} \bibinfo{volume}{1}
	(\bibinfo{year}{2013}) \bibinfo{pages}{011002}.
	\bibitem[{Jain et~al.(2011)Jain, Hautier, Ong, Moore, Fischer, Persson, and
		Ceder}]{jain2011formation}
	\bibinfo{author}{A.~Jain}, \bibinfo{author}{G.~Hautier}, \bibinfo{author}{S.~P.
		Ong}, \bibinfo{author}{C.~J. Moore}, \bibinfo{author}{C.~C. Fischer},
	\bibinfo{author}{K.~A. Persson}, \bibinfo{author}{G.~Ceder},
	\newblock \bibinfo{title}{Formation enthalpies by mixing {GGA and GGA+ U}
		calculations},
	\newblock \bibinfo{journal}{Physical Review B} \bibinfo{volume}{84}
	(\bibinfo{year}{2011}) \bibinfo{pages}{045115}.
	\bibitem[{Wang et~al.(2021)Wang, Kingsbury, McDermott, Horton, Jain, Ong,
		Dwaraknath, and Persson}]{wang2021framework}
	\bibinfo{author}{A.~Wang}, \bibinfo{author}{R.~Kingsbury},
	\bibinfo{author}{M.~McDermott}, \bibinfo{author}{M.~Horton},
	\bibinfo{author}{A.~Jain}, \bibinfo{author}{S.~P. Ong},
	\bibinfo{author}{S.~Dwaraknath}, \bibinfo{author}{K.~A. Persson},
	\newblock \bibinfo{title}{A framework for quantifying uncertainty in {DFT}
		energy corrections},
	\newblock \bibinfo{journal}{Scientific Reports} \bibinfo{volume}{11}
	(\bibinfo{year}{2021}) \bibinfo{pages}{15496}.
	\bibitem[{Aykol et~al.(2018)Aykol, Dwaraknath, Sun, and
		Persson}]{aykol2018thermodynamic}
	\bibinfo{author}{M.~Aykol}, \bibinfo{author}{S.~S. Dwaraknath},
	\bibinfo{author}{W.~Sun}, \bibinfo{author}{K.~A. Persson},
	\newblock \bibinfo{title}{Thermodynamic limit for synthesis of metastable
		inorganic materials},
	\newblock \bibinfo{journal}{Science Advances} \bibinfo{volume}{4}
	(\bibinfo{year}{2018}) \bibinfo{pages}{eaaq0148}.
	\bibitem[{Ong et~al.(2008)Ong, Wang, Kang, and Ceder}]{ong2008li}
	\bibinfo{author}{S.~P. Ong}, \bibinfo{author}{L.~Wang},
	\bibinfo{author}{B.~Kang}, \bibinfo{author}{G.~Ceder},
	\newblock \bibinfo{title}{Li--{Fe}--{P}--{O}$_2$ phase diagram from
		first-principles calculations},
	\newblock \bibinfo{journal}{Chemistry of Materials} \bibinfo{volume}{20}
	(\bibinfo{year}{2008}) \bibinfo{pages}{1798--1807}.
	\bibitem[{Barron and Klein(1965)}]{TBarron_1965}
	\bibinfo{author}{T.~H.~K. Barron}, \bibinfo{author}{M.~L. Klein},
	\newblock \bibinfo{title}{Second-order elastic constants of a solid under
		stress},
	\newblock \bibinfo{journal}{Proceedings of the Physical Society}
	\bibinfo{volume}{85} (\bibinfo{year}{1965}) \bibinfo{pages}{523}.
	\bibitem[{Larsen et~al.(2017)Larsen, Mortensen, Blomqvist, Castelli,
		Christensen, Dułak, Friis, Groves, Hammer, Hargus, Hermes, Jennings, Jensen,
		Kermode, Kitchin, Kolsbjerg, Kubal, Kaasbjerg, Lysgaard, Maronsson, Maxson,
		Olsen, Pastewka, Peterson, Rostgaard, Schiøtz, Schütt, Strange, Thygesen,
		Vegge, Vilhelmsen, Walter, Zeng, and Jacobsen}]{ase-paper}
	\bibinfo{author}{A.~H. Larsen}, \bibinfo{author}{J.~J. Mortensen},
	\bibinfo{author}{J.~Blomqvist}, \bibinfo{author}{I.~E. Castelli},
	\bibinfo{author}{R.~Christensen}, \bibinfo{author}{M.~Dułak},
	\bibinfo{author}{J.~Friis}, \bibinfo{author}{M.~N. Groves},
	\bibinfo{author}{B.~Hammer}, \bibinfo{author}{C.~Hargus},
	\bibinfo{author}{E.~D. Hermes}, \bibinfo{author}{P.~C. Jennings},
	\bibinfo{author}{P.~B. Jensen}, \bibinfo{author}{J.~Kermode},
	\bibinfo{author}{J.~R. Kitchin}, \bibinfo{author}{E.~L. Kolsbjerg},
	\bibinfo{author}{J.~Kubal}, \bibinfo{author}{K.~Kaasbjerg},
	\bibinfo{author}{S.~Lysgaard}, \bibinfo{author}{J.~B. Maronsson},
	\bibinfo{author}{T.~Maxson}, \bibinfo{author}{T.~Olsen},
	\bibinfo{author}{L.~Pastewka}, \bibinfo{author}{A.~Peterson},
	\bibinfo{author}{C.~Rostgaard}, \bibinfo{author}{J.~Schiøtz},
	\bibinfo{author}{O.~Schütt}, \bibinfo{author}{M.~Strange},
	\bibinfo{author}{K.~S. Thygesen}, \bibinfo{author}{T.~Vegge},
	\bibinfo{author}{L.~Vilhelmsen}, \bibinfo{author}{M.~Walter},
	\bibinfo{author}{Z.~Zeng}, \bibinfo{author}{K.~W. Jacobsen},
	\newblock \bibinfo{title}{The atomic simulation environment—a {Python}
		library for working with atoms},
	\newblock \bibinfo{journal}{Journal of Physics: Condensed Matter}
	\bibinfo{volume}{29} (\bibinfo{year}{2017}) \bibinfo{pages}{273002}.
	\bibitem[{Bahn and Jacobsen(2002)}]{ISI:000175131400009}
	\bibinfo{author}{S.~R. Bahn}, \bibinfo{author}{K.~W. Jacobsen},
	\newblock \bibinfo{title}{An object-oriented scripting interface to a legacy
		electronic structure code},
	\newblock \bibinfo{journal}{Comput. Sci. Eng.} \bibinfo{volume}{4}
	(\bibinfo{year}{2002}) \bibinfo{pages}{56--66}.
	\bibitem[{Ong et~al.(2013)Ong, Richards, Jain, Hautier, Kocher, Cholia, Gunter,
		Chevrier, Persson, and Ceder}]{ong2013python}
	\bibinfo{author}{S.~P. Ong}, \bibinfo{author}{W.~D. Richards},
	\bibinfo{author}{A.~Jain}, \bibinfo{author}{G.~Hautier},
	\bibinfo{author}{M.~Kocher}, \bibinfo{author}{S.~Cholia},
	\bibinfo{author}{D.~Gunter}, \bibinfo{author}{V.~L. Chevrier},
	\bibinfo{author}{K.~A. Persson}, \bibinfo{author}{G.~Ceder},
	\newblock \bibinfo{title}{Python materials genomics (pymatgen): A robust,
		open-source {Python} library for materials analysis},
	\newblock \bibinfo{journal}{Computational Materials Science}
	\bibinfo{volume}{68} (\bibinfo{year}{2013}) \bibinfo{pages}{314--319}.
	\bibitem[{Bitzek et~al.(2006)Bitzek, Koskinen, G{\"a}hler, Moseler, and
		Gumbsch}]{bitzek2006structural}
	\bibinfo{author}{E.~Bitzek}, \bibinfo{author}{P.~Koskinen},
	\bibinfo{author}{F.~G{\"a}hler}, \bibinfo{author}{M.~Moseler},
	\bibinfo{author}{P.~Gumbsch},
	\newblock \bibinfo{title}{Structural relaxation made simple},
	\newblock \bibinfo{journal}{Physical Review Letters} \bibinfo{volume}{97}
	(\bibinfo{year}{2006}) \bibinfo{pages}{170201}.
	\bibitem[{Liu et~al.(2024)Liu, Liu, Riebesell, Qi, Ong, and Ko}]{matcalc2024}
	\bibinfo{author}{R.~Liu}, \bibinfo{author}{E.~Liu},
	\bibinfo{author}{J.~Riebesell}, \bibinfo{author}{J.~Qi},
	\bibinfo{author}{S.~P. Ong}, \bibinfo{author}{T.~W. Ko},
	\bibinfo{title}{Matcalc: A python library for calculating materials
		properties from the potential energy surface (pes)}, \bibinfo{year}{2024}.
	\bibitem[{Singh et~al.(2021)Singh, Lang, Dovale-Farelo, Herath, Tavadze,
		Coudert, and Romero}]{singh2021mechelastic}
	\bibinfo{author}{S.~Singh}, \bibinfo{author}{L.~Lang},
	\bibinfo{author}{V.~Dovale-Farelo}, \bibinfo{author}{U.~Herath},
	\bibinfo{author}{P.~Tavadze}, \bibinfo{author}{F.-X. Coudert},
	\bibinfo{author}{A.~H. Romero},
	\newblock \bibinfo{title}{Mechelastic: A {Python} library for analysis of
		mechanical and elastic properties of bulk and 2d materials},
	\newblock \bibinfo{journal}{Computer Physics Communications}
	\bibinfo{volume}{267} (\bibinfo{year}{2021}) \bibinfo{pages}{108068}.
	\bibitem[{Liu et~al.(2025)Liu, Zeng, Wang, and Zhao}]{Fine-Tuning}
	\bibinfo{author}{X.~Liu}, \bibinfo{author}{K.~Zeng}, \bibinfo{author}{Y.~Wang},
	\bibinfo{author}{T.~Zhao},
	\newblock \bibinfo{title}{A study on the fine-tuning performance of universal
		machine-learned interatomic potentials {(uMLIPs)}},
	\newblock \bibinfo{journal}{arXiv:2506.07401}  (\bibinfo{year}{2025}).
	\bibitem[{Zhang et~al.(2020)Zhang, Wang, Chen, Zeng, Zhang, Wang
		et~al.}]{zhang2020dp}
	\bibinfo{author}{Y.~Zhang}, \bibinfo{author}{H.~Wang},
	\bibinfo{author}{W.~Chen}, \bibinfo{author}{J.~Zeng},
	\bibinfo{author}{L.~Zhang}, \bibinfo{author}{H.~Wang}, et~al.,
	\newblock \bibinfo{title}{Dp-gen: A concurrent learning platform for the
		generation of reliable deep learning based potential energy models},
	\newblock \bibinfo{journal}{Computer Physics Communications}
	\bibinfo{volume}{253} (\bibinfo{year}{2020}) \bibinfo{pages}{107206}.
	
\end{thebibliography}
	
\end{document}